\documentclass{article}
\usepackage{hyperref}
\usepackage{amssymb}
\usepackage{color}
\definecolor{MyDarkBlue}{rgb}{0.1,0,0.80} 
\definecolor{MyRed}{rgb}{1,0,0}
\definecolor{MyYellow}{rgb}{1,1,0.8}
\hypersetup{colorlinks=true, citecolor=blue, linkcolor= MyDarkBlue}

\newcommand{\nat}{Nature}
\newcommand{\apjl}{ApJ}
\newcommand{\apj}{ApJ}
\newcommand{\apjs}{ApJS}
\newcommand{\mnras}{MNRAS}

\newcommand{\aapr}{A{\&}ARv}

\newcommand{\prd}{PRD}

\title{The maximum dipole moments of neutron stars implied by the Barrow-Gibbons conjecture}

\author{\href{http://www.fizik.itu.edu.tr/eksiy/}{Kaz{\i}m Yavuz Ek\c{s}i}\footnote{\href{mailto:eksi@itu.edu.tr}{eksi@itu.edu.tr}}. \\
\href{http://www.itu.edu.tr/}{\.Istanbul Technical University}, \href{http://www.itu.edu.tr/akademi/fakulteler/fen-edebiyat-fakultesi}{Faculty of Sciences and Letters}, \\
\href{http://www.fizik.itu.edu.tr/tr/}{Department of Physics}, 34469 Maslak, \.Istanbul
}

\begin{document}

\maketitle

\begin{abstract}
The maximum magnetic moment to angular momentum conjecture, recently posed by Barrow \& Gibbons, is in tension with the generally accepted parameters of magnetars. 
According to this conjecture the dimensionless Schuster-Wilson-Blacket number, $c\mu/J\sqrt{G}$ where $\mu$ is the magnetic moment and $J$ is the angular momentum, should be of order unity.
A resolution is to assume that not only the low-magnetic field magnetars, but all magnetars have their super-critical fields in the higher multipoles. 
This requires the presence of external torques other than the magnetic dipole torque, such as from fallback disks or winds, on these systems to yield the measured high spin-down rates.
\end{abstract}

\section{Introduction}

Magnetars are neutron stars showing sporadic bursts with luminosity $L\sim 10^{41}$~erg~s$^{-1}$ which well exceed their Eddington luminosity $L_{\rm E} \sim 10^{38}$~erg~s$^{-1}$ \cite{mer08}. 
Three of them also showed  giant bursts of $\sim 10^{44}-10^{46}$~erg~s$^{-1}$ \cite{maz+79,hur+99,hur+05}. 
The magnetar model \cite{dun92,tho93,tho95,tho96} explains these phenomena by the existence of magnetic fields beyond the quantum critical value,
\begin{equation}
B_{\rm c} \equiv \frac{m_{\rm e}^2 c^3}{\hbar e} = 4.4\times 10^{13}~{\rm G},
\end{equation}
at which the cyclotron energy of the electron, $\hbar \omega_{\rm c}$ is equal to its rest mass energy, $m_e c^2$. 
The rapid spin-down rate of magnetars \cite{kou98} imply magnetic dipole moments of order $\mu \sim 10^{33}~{\rm G~cm^3}$
assuming these objects spin-down under magnetic dipole radiation torque
\begin{equation}
I \frac{d\Omega}{dt} = -\frac{2\mu^2 \sin^2 \alpha }{3c^3}\Omega^3
\label{mdr}
\end{equation}
where $I$ is the moment of inertia and $\Omega $ is the angular velocity of  the star and $\alpha$ is the angle between the rotation and magnetic axis.

According to the magnetar model even stronger fields of $B\sim 10^{16}-10^{17}~{\rm G}$ exist inside magnetars. This is still less than
the maximum critical magnetic field, $B_{\rm c} \sim 10^{18}~{\rm G}$ set by balance of the gravitational potential energy, $\sim GM^2/R$, 
with the magnetic energy $\sim B^2 R^3/6$, 
the so called Chandrasekhar-Fermi limit \cite{cha53} for a self-gravitating object.

According to maximum magnetic moment to angular momentum conjecture recently posed by Barrow
and Gibbons \cite{bar17} the dimensionless Schuster-Wilson-Blackett number \cite{sch12,wil23,bla47}
\begin{equation}
\theta \equiv \frac{c\mu }{J\sqrt{G}},
\label{swr}
\end{equation}%
where $\mu $ is the magnetic moment and $J$ is the angular momentum, is
bounded from above by a number of order unity, $\theta_{\rm c}$. 
We show here that this limits the magnetic dipole field of neutron stars more tightly than the Chandrasekhar-Fermi limit.
This results with a much tighter bound on the \textit{dipole} magnetic moment of neutron stars and is in strong tension with the inferred values of magnetic fields for magnetars.

\section{Barrow-Gibbons limit for neutron stars}

The angular momentum of a neutron star is $J=I\Omega$ where $I=\frac{2}{5}MR^{2}\sim 10^{45}~{\rm g~cm^2}$. Using these in \autoref{swr} one finds
\begin{equation}
\theta =\frac{5 c}{2 M R^2 \sqrt{G}} \frac{\mu}{\Omega} <\theta_{\rm c}
\end{equation}
where $\theta_{\rm c}$ is assumed to be of order unity by Barrow
and Gibbons \cite{bar17}.
Scaling this with appropriate values for magnetars we obtain
\begin{equation}
\theta=23.1\left( \frac{\mu }{10^{33}~{\rm G~cm^{3}}}\right) \left( \frac{P}{1~{\rm s}}\right)
\left( \frac{M}{M_{\odot}}\right)^{-1}\left( \frac{R}{10^{6}~{\rm cm}}\right)^{-2}
\end{equation}
where $P=2\pi /\Omega $ is the spin period of the star and ranges between $P=2-12$~s for magnetars\footnote{See the magnetar catalogue \url{http://www.physics.mcgill.ca/~pulsar/magnetar/main.html}} \cite{ola14}.
This result is in strong tension with the Barrow-Gibbons conjecture stating that $\theta_{\rm c} \sim 1$. The problem becomes even more acute if one considers that some of the magnetars have periods $P\sim 10$~s and $\mu \sim 2 \times 10^{33}$~G~cm$^3$. 
The dipole moment as inferred from \autoref{mdr} is
\begin{equation}
\mu =1.1 \times 10^{32}\left( \frac{P}{1~{\rm s}}\frac{\dot{P}}{10^{-11}}\right) ^{1/2}
\end{equation}
For the parameters of magnetar 1RXS~J170849.0--400910, e.g.\ with $P=11~{\rm s}$ and $\dot{P}=1.95\times 10^{-11}$~s~s$^{-1}$ \cite{dib14}
this implies $\mu = 5 \times 10^{32}~{\rm G~cm^3}$.
Accordingly, this gives $\theta_{\rm c} \gtrsim 90$ for this object for $M=1.4~M_{\odot}$. The strongest magnetic moment inferred for a magnetar is that of SGR~1806--20 which has a period of $P=7.55$~s and a period derivative of $\dot{P} \simeq 4.95\times 10^{-10}~{\rm s~s^{-1}}$ \cite{woo+07} implying 
$\mu = 2\times 10^{33}~{\rm G~cm^3}$ and $\theta_{\rm c}\gtrsim 350$, three orders of magnitude larger than the Barrow-Gibbons conjecture requires.

\section{Discussion}

We have shown that the recently proposed Barrow-Gibbons conjecture of the ratio of magnetic moment to angular momentum for a rotating body \cite{bar17} that the dimensionless Schuster-Wilson-Blackett number \cite{sch12,wil23,bla47} should be of order unity tightly limits the magnetic dipole moment of neutron stars as compared to the Chandrasekhar-Fermi limit \cite{cha53}. 
This has strong implications for highest magnetic field objects in the Universe, i.e.\ magnetars.
The proposed dipole fields of magnetars \cite{dun92,tho96} with magnetic moments of order $\mu \sim 10^{33}$~G~cm$^3$ leads to values of Schuster-Wilson-Blackett number of order $\sim 100$ for magnetars implying that $\theta_{\rm c}$ should exceed 350.
This suggests either of the following possibilities:
\begin{itemize}
\item The Barrow-Gibbons conjecture is not realized in nature or is not sufficiently general to apply macroscopic objects like neutron stars.
\item The critical Schuster-Wilson-Blackett number is not of unity but much larger i.e.\ $\theta_{\rm c} \gtrsim 10^3$, due to some unknown dimensionless factors like $(4  \pi)^3$ etc.
\item The magnetars do not have such strong \textit{dipole} fields.
\end{itemize}

Here we elaborate on the final possibility assuming Barrow-Gibbons conjecture applies to neutron stars. The magnetar bursting activity may not require the existence of ultr-strong dipole moments 
if the super-critical fields leading to the bursts could be in the higher multipoles \cite{tho95,gav02,eks03,ert03,tie+13}. 
The recently identified low-magnetic field magnetars \cite{rea+10,rea+12} indeed indicate that the dipole field is not the critical parameter rendering a neutron star to be a magnetar. 
In order to comply with the Barrow-Gibbons conjecture, it is reasonable to suggest that not only low-B magnetars, but all magnetars have magnetic dipole moments 
If, however, one assumes that \textit{all} magnetars have smaller dipole fields it would not be possible to explain the rapid-spin-down of these objects by the magnetic dipole torque alone. 
This would require the assistance of other torques contributing to the spin-down. 
The presence of propeller torques from putative supernova fallback disks \cite{alp01,cha+00} was suggested to obviate such strong dipole fields for such young neutron stars. 
Infrared emission possibly originating from such a disk was first detected from 
two magnetars \cite{wan+06,kap09} though the authors of these papers presumed these disks were inactive. 
Action of the disk torque requires that such disks are active and such a model was provided by the authors of Refs.\cite{ert+07,ert+09}.
Yet another possibility that could provide additional torque could be the ejection of relativistic particles from magnetars \cite{tho+04,buc+07}; this, however, is not proposed as a continuous process, but suggested to prevail at the youngest stages.

Recently discovered pulsating ultra-luminous X-ray sources (PULXs) \cite{bac+14,isr+17a,isr+17b,fur+16} 
are also suggested to  have very strong dipole magnetic fields $B \sim \mbox{ a few } 10^{13}~{\rm G}$ \cite{eks+15}. 
The latter authors suggested that---as with the case of magnetars---even higher fields could exist in the higher multipoles 
which was recently supported by the observational results \cite{isr+17a,isr+17b}. 
If this is the case then this could imply that strong quadrupole magnetic fields
are very common among young neutron stars either isolated or in binaries. Distributing the magnetic flux to higher multipoles could be Nature's way of complying with Barrow-Gibbons conjecture.


\end{document}